\documentclass[12pt]{iopart}
\usepackage{epsfig}
\usepackage{graphicx}

\begin{document}

\title{Driven particle in a cloud of mobile impurities}

\author{Lasse Laurson and Mikko J. Alava}

\address{Laboratory of Physics, Helsinki University of Technology,
FIN-02015 HUT, Finland}

\eads{\mailto{lla@fyslab.hut.fi}, \mailto{mja@fyslab.hut.fi}}

\begin{abstract}
The dynamics of a test particle interacting with diffusing
impurities in one dimension is investigated analytically and
numerically. In the absence of an applied external force, the
dynamics of the particle can be
characterized by a distribution of monotonic excursions
$\Delta x$, which scales as a power law with an exponent
$\tau_{\Delta x} = 4/3$. When the particle is driven at a slow
constant velocity, there is again a power law
distribution for the monotonic changes of the force $\Delta F$, which is
characterized by a similar exponent $\tau_{\Delta F}=4/3$. These
results can be understood from the theory of random walks.
\end{abstract}


\maketitle

\section{Introduction}\label{intro}

The interaction of driven particles, flexible lines and membranes etc.
with disorder is an important topic in condensed matter physics 
\cite{NAT-83,HUS-85,KAR-87}. Usually, this disorder is taken to be 
{\it quenched}, or frozen, such that its properties do not change 
within the relevant time scales. However, under certain conditions, 
this changes as in the case of the diffusion of solute atoms in 
metallic alloys \cite{CAH-62,BLA-99} or oxygen vacancies in 
superconductors \cite{CHU-98}. The mobile impurities play an important 
role in the dynamics of such systems, as evidenced for example by the 
Portevin-Le Chatelier (PLC) effect in solid solutions \cite{POR-23}. 
There, within a certain range of temperatures and applied strain rates, 
the dynamic interaction of lattice dislocations and diffusing solute 
atoms result in phenomena such as negative strain rate sensitivity of 
the flow stress, giving rise to macroscopic serrations in the stress
strain curve and strain localization in the form of bands of activity 
of various types \cite{ANA-99,LEB-95,HAH-02}.

Here, we consider the simple test problem of a single particle 
interacting with a cloud of diffusing impurities, with the dynamics 
constrained in one dimension (a line). We restrict ourselves to the 
region of the parameter space in which the impurities have a vanishingly 
small probability to escape from the vicinity of the particle. 
Despite its apparent simplicity, such a system exhibits rich dynamics, 
but at the same time has features that make the problem analytically 
tractable. In the absence of external forces, we consider the
statistics of {\it monotonic excursions} $\Delta x$ of the particle, i.e. 
the distances the particle moves to a particular direction (here 
``left'' or ``right'' along the one-dimensional line) without changing 
direction, see also figure \ref{fig:trajectories}. We find that these 
obey power law distributions 
$P(\Delta x) \sim (\Delta x)^{-\tau_{\Delta x}}f_c(\Delta x/\Delta x_0)$
with the exponent $\tau_{\Delta x}=4/3$. The same is true for 
the monotonic changes of the external force $\Delta F$ when the particle 
is driven with a slow constant velocity. 
This paper is organized as follows: In the next section, we consider the
interaction of a particle with a single mobile impurity, in the
absence of external forces. Then we generalize this to the case with
more impurity particles. In section \ref{driven}, the effect of external drive
is studied. Finally, section \ref{concl} finishes the paper with conclusions.

\section{A particle interacting with mobile impurities}\label{nondriven}

\subsection{Single impurity}

As a starting point of our analysis, we consider the dynamics of a single
particle interacting with one diffusing impurity particle. The equations
of motion for the system are
\begin{eqnarray}
\label{eq:motion}
\mu \partial_t x & = & f(x-x_s) \nonumber\\
\partial_t x_s &=& -f(x-x_s) + \eta,
\end{eqnarray}
where $x$ and $x_s$ are the positions of the particle and the impurity
particle, respectively. $f(z)$ is the interaction force between the
particle and the impurity particle, $\mu$ defines the relative mobilities
of the impurity and the particle and $\eta$ is Gaussian white noise
with standard deviation $\delta \eta$ mimicking the effect of temperature.

The dynamics of the particle can be analyzed by considering the
stochastic process for the velocity $\partial_t x$. By differentiating
the equation of motion of the particle with respect to time and using
the equation of motion for $x_s$, one obtains
\begin{equation}
\label{eq:x1}
\mu \partial_t^2 x= \partial_z f(z) \left[(1+\mu)\partial_t x - \eta \right].
\end{equation}
Close to $z=x-x_s=0$, the force $f(z)$ can be taken to be linear in
$z$ and thus the derivative of the force can be approximated by a
constant, $\partial_z f(z) \approx -C$, with $C>0$. With $\lambda =
C(1+\mu)/\mu$ and $\xi=-(C/\mu) \eta$, equation (\ref{eq:x1}) can then be
rewritten in the form of an Ornstein-Uhlenbeck process for $\partial_t x$,
\begin{equation}
\label{eq:x2}
\partial_t^2 x = -\lambda \partial_t x + \xi.
\end{equation}
For the process $\partial_t x$, equation (\ref{eq:x2}) describes Brownian
motion pushed toward the origin by a linear damping term. This problem
has been considered e.g. in \cite{COL-04}, and the scaling
exponents are known. In particular, the first return times $T$ to
origin of $\partial_t x$ have a probability distribution scaling as
\begin{equation}
\label{eq:pt}
P(T) \sim T^{-\tau_T} f_c\left(\frac{T}{T_0}\right), 
\end{equation}
with $\tau_T=3/2$ and the cut-off
scale $T_0 \sim 1/\lambda$. Similarly, the average shape of an excursion,
$\langle \partial_t x(t)\rangle_T$, scales for $t, T-t \ll 1/\lambda$ as
$\langle \partial_t x(t)\rangle_T = T^{\gamma-1}f_{shape}(t/T)$,
with $\gamma=3/2$. Then, by using the scaling relation
$\gamma=(\tau_T-1)/(\tau_{\Delta x}-1)$ \cite{LUB-04}, one obtains the
distribution of the lengths $\Delta x = \int_0^T \partial_t x dt$ of
monotonic excursions of the particle,
\begin{equation}
\label{eq:dist}
P(\Delta x) \sim (\Delta x)^{-\tau_{\Delta x}}
f_c\left(\frac{\Delta x}{\Delta x_0}\right),
\end{equation}
with $\tau_{\Delta x} = 4/3$. The cut-off scaling can be found as follows:
The cut-off of the first return time distribution of equation (\ref{eq:x2})
is given by $T_0 \sim 1/\lambda$ \cite{COL-04}. Here, both $\lambda$ and
$\xi$ in equation (\ref{eq:x2}) depend on $C$ and $\mu$. Due to the relation
$\Delta x_0 \sim T_0^{\gamma}$ with $\gamma=3/2$, $\Delta x_0$ is expected 
to scale like
\begin{equation}
\label{eq:cutoff1}
\Delta x_0 = \left(\frac{1}{\lambda}\right)^{3/2} \cdot \delta \xi =
\sqrt{\frac{\mu}{C}}\frac{\delta \eta}{(1+\mu)^{3/2}}.
\end{equation}

\begin{figure}[!h]
\begin{center}
\includegraphics[width = 9cm]{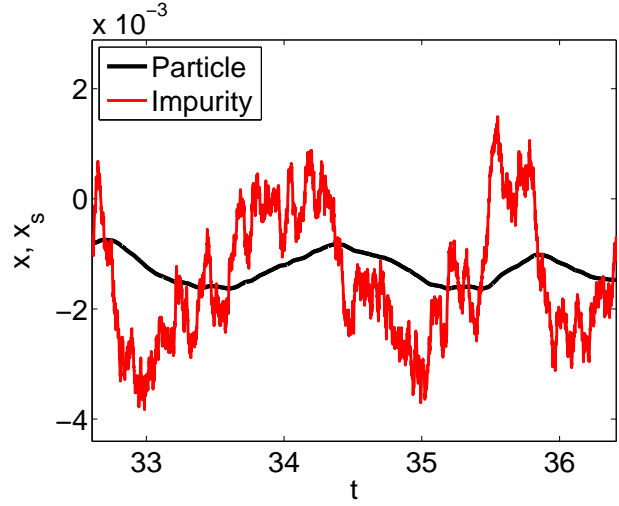}
\end{center}
\caption{An example of the trajectories of the particle and a single
    impurity, in the absence of external forces. Notice that the monotonic
    excursions of the particle correspond to motion of the particle during
    time intervals the impurity spends on a given side of the particle. 
    Parameters of the simulation: $\mu=A=l=1.0$, $\delta \eta=0.1$.}
\label{fig:trajectories}
\end{figure}

\begin{figure}[!h]
\begin{center}
\includegraphics[width = 9cm]{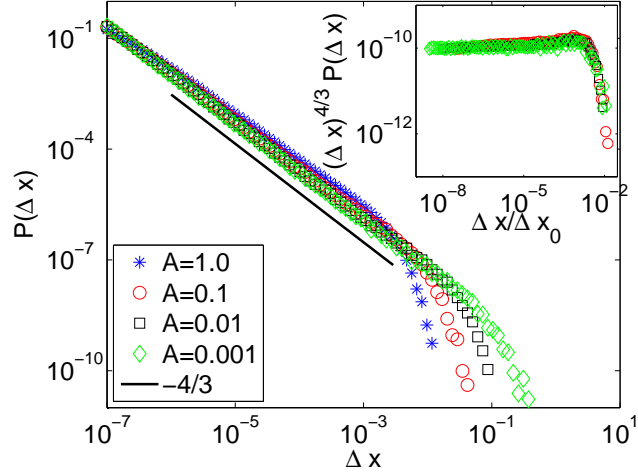}
\end{center}
\caption{Main Figure: The probability distributions of monotonic
excursions of the particle interacting with a single impurity,
for various strengths $A$ of the interaction force. For sufficiently
low $A$ values, the distribution scales with the exponent
$\tau_{\Delta x} = 4/3$, indicated by the solid line. Inset: A scaling
plot of the distributions. The cut-off is observed to scale as
$\Delta x_0 \sim \sqrt{1/A}$, in agreement with equation (\ref{eq:cutoff1}).
Other parameters: $l=1$, $\mu=1$ and $\delta \eta=0.1$.}
\label{fig:onesolute}
\end{figure}

We check this result in numerical simulations, where we
take for simplicity $f(z)$ to be of the form 
$f(z) = -A z \exp{(-(1/2)(z/l)^2)}$, corresponding to 
$\partial_z f(z)|_{z=0}=-A$, i.e. $C=A$. We integrate the equations
of motion (\ref{eq:motion}) with the Euler algorithm. The strength 
of the thermal noise was chosen to be sufficiently weak such that 
the impurity cannot escape from the neighborhood of the particle. Figure 
\ref{fig:trajectories} shows an example of the trajectories of the 
particle and the impurity. Figure \ref{fig:onesolute} 
displays the distribution of $\Delta x$ for various values of $A$. 
For a weak enough interaction strength A, the distributions display 
the expected scaling with $\tau_{\Delta x} = 4/3$. The cut-off is 
found to scale as $\Delta x_0 \sim A^{-1/2}$, in agreement with our 
results above.

\begin{figure}[!h]
\begin{center}
\includegraphics[width = 9cm]{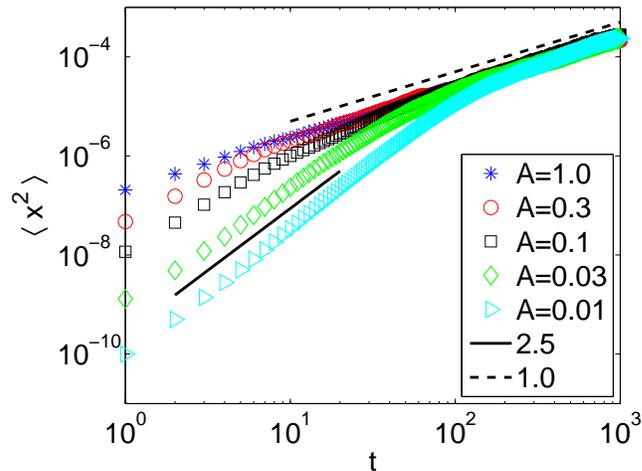}
\end{center}
\caption{The mean square displacement of the particle interacting
with a single impurity in the absence of external forces, for 
different strengths $A$ of the particle-impurity interaction.
The early times are characterized by superballistic motion with 
$\langle x^2 \rangle(t) \sim t^{5/2}$ (solid line), while for
longer times a cross-over to diffusive dynamics is observed (dashed
line). Parameters of the simulation: $\mu=l=1.0$, $\delta 
\eta=0.1$.}
\label{fig:diffusion}
\end{figure}

This implies that the motion of the particle is reminiscent
of {\it truncated} Levy flight, with a step length distribution 
given by equation (\ref{eq:dist}). However, the steps are not
instantaneous - their durations $T$ exhibit power law scaling 
as well, equation (\ref{eq:pt}). By interpreting these step durations 
as waiting times between instantaneous steps, one would obtain
for early times the scaling
$\langle x^2 \rangle(t) \sim t^{2 (\tau_T-1)/(\tau_{\Delta x}-1)}
= t^{2\gamma} \sim t^{3}$.
However, here $\Delta x$ and $T$ are not independent (due to
the relation $\langle \Delta x \rangle \sim T^{\gamma}$). As the 
early time behavior is dominated by a single large step, one must 
consider instead the effect of a single step given its duration,
\begin{equation}
\langle x^2 \rangle(t) \sim \int_0^t [\Delta x(T)]^2 P(T) dT
=\int_0^t T^{2\gamma-\tau_T} dT 
\sim  t^{2\gamma-\tau_T+1},
\end{equation}
corresponding to $\langle x^2 \rangle(t) \sim t^{5/2}$.
Due to the truncated nature of the Levy flight, one expects a
cross-over to diffusive behavior with $\langle x^2 \rangle(t) \sim t$
for long times. This is verified in figure \ref{fig:diffusion}.

\subsection{Several impurities}

\begin{figure}[!t]
\begin{center}
\includegraphics[width = 9cm]{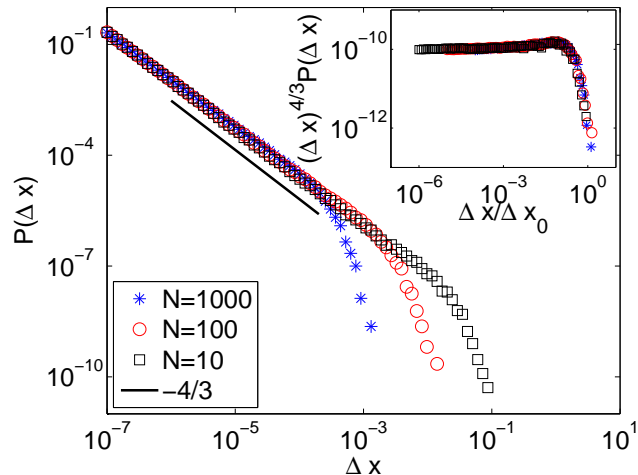}
\end{center}
\caption{Main Figure: The probability distributions of monotonic
excursions of the particle interacting with $N$ impurities.
The distribution scales with the exponent $\tau_{\Delta x} = 4/3$,
indicated by the solid line. Inset: A scaling
plot of the distributions. The cut-off is observed to scale as
$\Delta x_0 \sim 1/N$, in agreement with equation (\ref{eq:cutoff2}).
Parameters of the simulation were
$A=0.01$,$l=1$, $\mu=1$ and $\delta \eta=0.1$.}
\label{fig:manysolutes}
\end{figure}

The case of a fixed number $N>1$ of impurity particles is a
straightforward generalization of that presented in the previous
subsection. The equations of motion become
\begin{eqnarray}
\mu \partial_t x & = & \sum_{i} f(x-x_{s,i})  \nonumber\\
\partial_t x_{s,i} &=& -f(x-x_{s,i}) + \eta_i.
\end{eqnarray}
From these equations, with the same procedure as above, one obtains
an equation of the form of equation (\ref{eq:x2}) by setting
$\lambda = C(N+\mu)/\mu$ and $\xi = (C/\mu)\sum_i\eta_i$. Thus,
the same scaling, i.e. $P(\Delta x) \sim (\Delta x)^{-4/3}$ is
expected. One should notice, however, that the cut-off scale
$\Delta x_0$ is getting smaller with increasing $N$. With similar
arguments as above, one finds that
\begin{equation}
\label{eq:cutoff2}
\Delta x_0 = \left(\frac{1}{\lambda}\right)^{3/2} \cdot \delta \xi =
\sqrt{\frac{\mu N}{C}}\frac{\delta \eta}{(N+\mu)^{3/2}}.
\end{equation}
This is verified by the numerical results presented in figure
\ref{fig:manysolutes}.

\section{Constant velocity drive}\label{driven}

Next we proceed to study the effect of a weak external force $F$ on
the dynamics of the particle. In this context a {\it constant
velocity} drive is perhaps the more interesting form of driving as a
small {\it constant force} with $\partial_t F = 0$ does not change
the dynamics from the non-driven case: equation (\ref{eq:x1}) remains 
the same even if a constant force term is introduced in equation 
(\ref{eq:motion}). In particular, we consider a particle driven by 
a force given by $F=K(Vt-x)$, where $V$ is the driving velocity 
and $K$ is a spring constant characterizing the response of the 
driving mechanism. The equations of motion read
\begin{eqnarray}
\mu \partial_t x & = & \sum_{i} f(x-x_{s,i}) + F \nonumber\\
\partial_t x_{s,i} &=& -f(x-x_{s,i}) + \eta_i.
\end{eqnarray}
In systems like this driven with a constant velocity, the
interesting quantity is the statistics of the external force
fluctuations. To this end, we consider the stochastic process
$\partial_t F$. With a similar approach as above, one can
write
\begin{eqnarray}
\label{eq:f1}
\partial_t^2 F & = & -K\partial_t^2 x \nonumber \\
& = & -\left[\frac{K}{\mu} + \frac{C}{\mu}(N+\mu)\right]\partial_t F
+ \frac{KC}{\mu}\sum_i \eta_i \nonumber \\
& & + \frac{KC}{\mu}\left[V(N+\mu)-F\right],
\end{eqnarray}
where the relation $\partial_t x = V-\partial_t F/K$ has been used.
In the steady state, the last term in equation (\ref{eq:f1}) has a 
zero mean, as one can write for the average steady state force
$F_s=\langle \mu \partial_t x - \sum_i f(x-x_{s,i})\rangle = \mu V + N f_s$,
where $f_s$ is the magnitude of the average retarding force acting
on the particle due to a single impurity. In the steady state the
condition $\langle \partial_t x_{s,i}\rangle = f_s = V$ holds, and thus
$\langle V(N+\mu)-F \rangle=0$. Assuming that the fluctuations
$\delta [V(N+\mu)-F] = \delta F$ are small compared to those of the white
noise term in equation (\ref{eq:f1}), i.e. 
$\delta F \ll \sqrt{N}\delta \eta$, equation (\ref{eq:f1}) can be 
approximately written as
\begin{equation}
\partial_t^2 F = -\left[\frac{K}{\mu} + \frac{C}{\mu}(N+\mu)\right]\partial_t F
+ \frac{KC}{\mu}\sum_i \eta_i,
\end{equation}
which is again of the same form as equation (\ref{eq:x2}). Thus, the 
monotonic changes of the external force $\Delta F = \int_0^T \partial_t F dt$ 
are expected to be distributed according to a power law $P(\Delta F) =
(\Delta F)^{-\tau_{\Delta F}} f_c(\Delta F/\Delta F_0)$, with the exponent
$\tau_{\Delta F} = 4/3$ and the cut-off scale $\Delta F_0$ scaling as
\begin{equation}
\label{eq:cutoff}
\Delta F_0 = \frac{KC \sqrt{\mu N} \delta \eta}{\left[K + C(N+\mu)\right]^{3/2}}.
\end{equation}
Notice that the condition $\delta F \ll \sqrt{N}\delta \eta$ implies that
\begin{equation}
\label{eq:cond}
\frac{KC\sqrt{\mu}}{\left[ K+C(N+\mu)\right]^{3/2}} \ll 1.
\end{equation}
For most of the relevant parameter values condition (\ref{eq:cond}) 
is fulfilled, only for $KC \gg 1$ this is not the case.

\begin{figure}[!t]
\begin{center}
\includegraphics[width = 9cm]{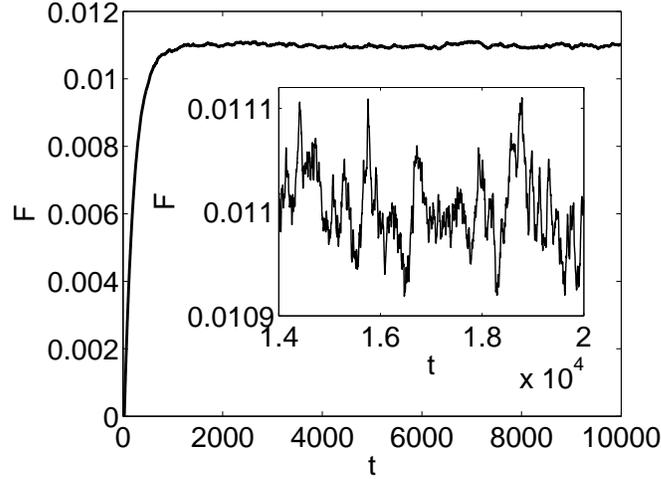}
\end{center}
\caption{An example of the behaviour of the force as a function of
  time. The inset shows a magnification of a part of the signal in
  the steady state. Parameters of the
  simulation: $\mu=l=1.0$, $A=0.01$, $N=10$, $K=0.1$,
  $\delta \eta=0.1$ and $V=0.001$.}
\label{fig:force}
\end{figure}

\begin{figure}[!h]
\begin{center}
\includegraphics[width = 9cm]{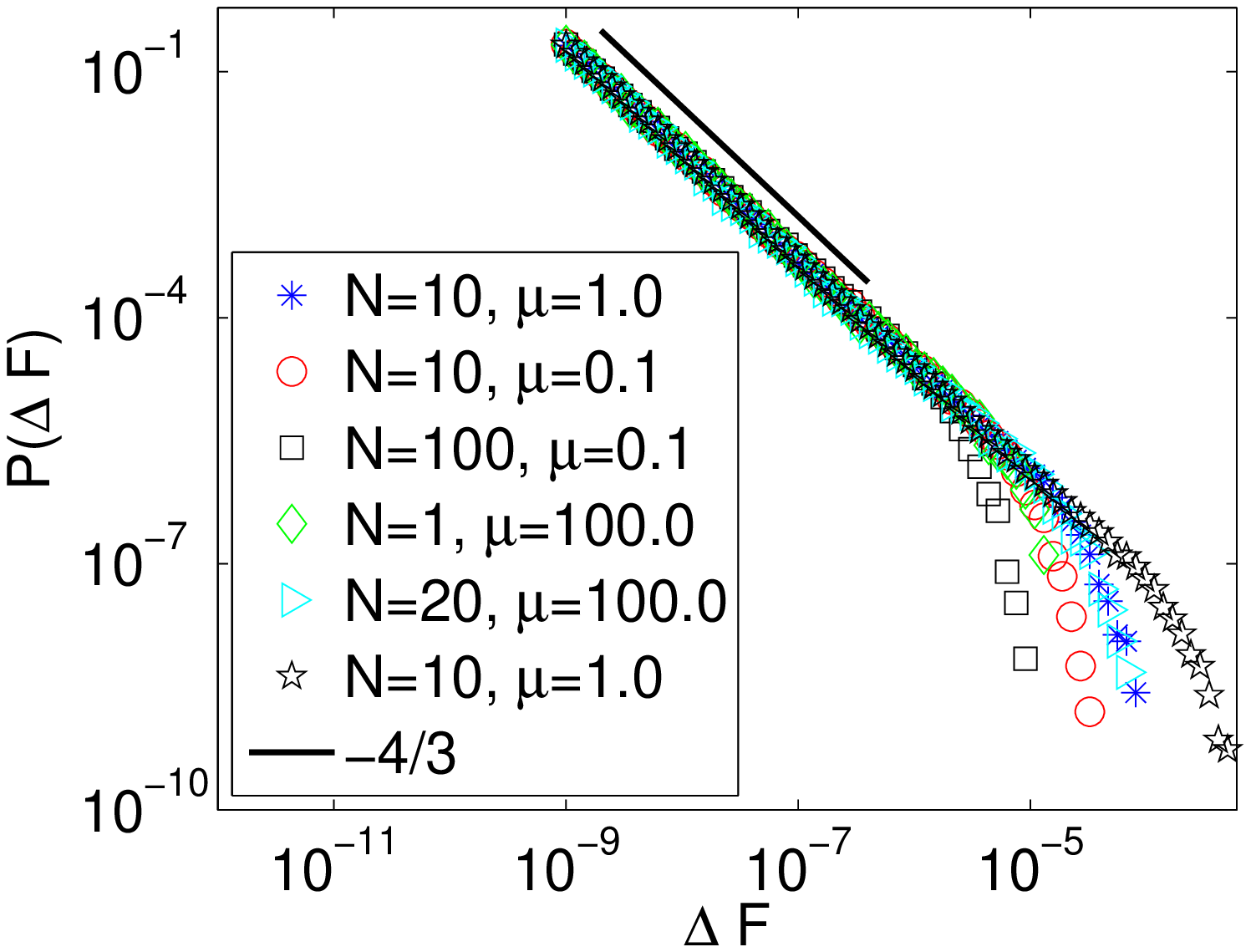} \\
\includegraphics[width = 9cm]{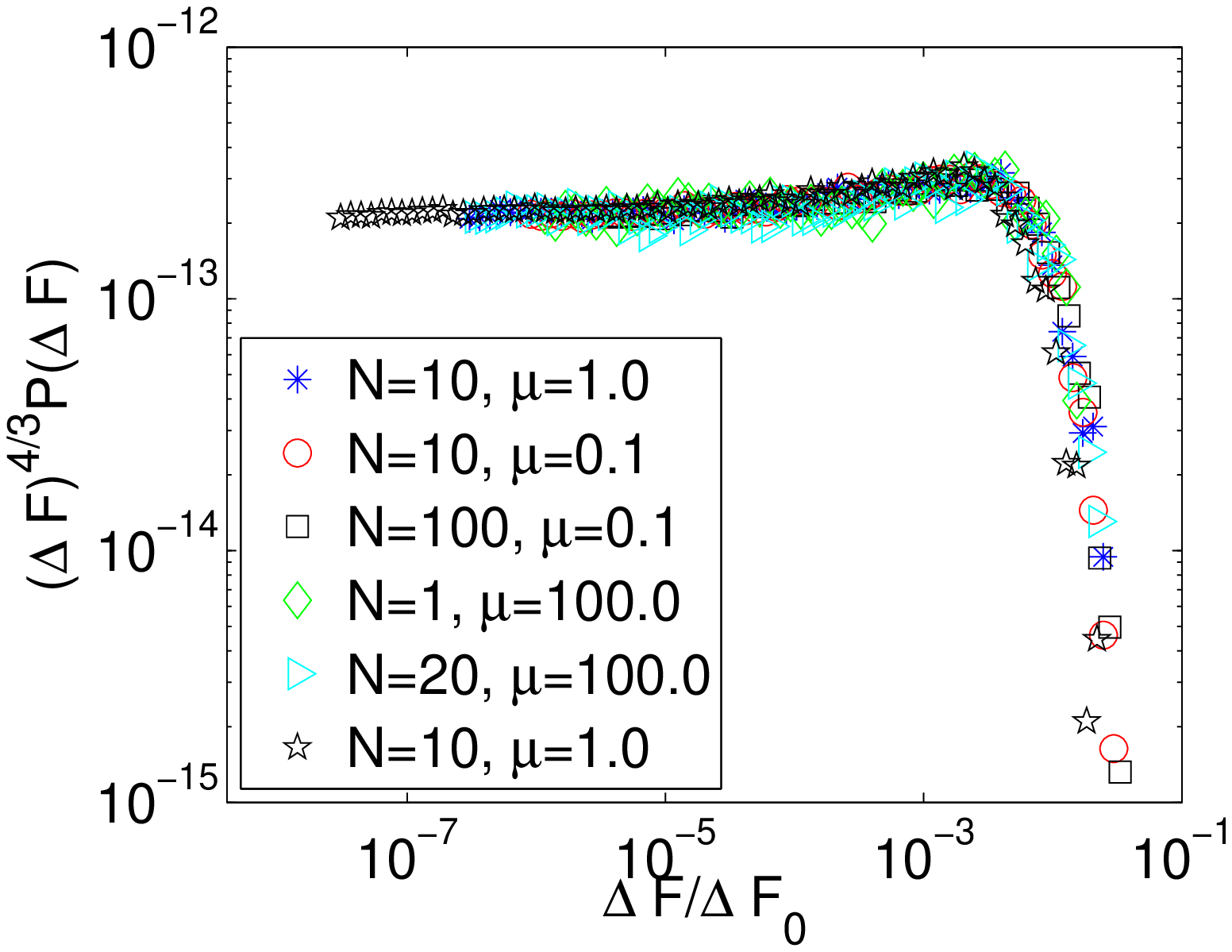}
\end{center}
\caption{Distributions of the monotonic changes of the external
force $\Delta F$ (upper panel) for various values of the parameters.
The lower panel shows a data collapse of the distributions, with
$\Delta F_0$ computed from equation (\ref{eq:cutoff}). All other cases
have $\delta\eta=0.1$, while the data shown with black stars has
$\delta\eta=1.0$. Other parameters of the simulations: $K=0.1$,
$A=0.01$ and $V=0.001$.}
\label{fig:forcedists}
\end{figure}

Next we check these predictions numerically for different values
of the parameters satisfying the condition (\ref{eq:cond}).
Figure \ref{fig:force} shows an example of the behaviour of the force as
a function of time. After an initial transient, the system reaches a
steady state in which the force fluctuates around a constant average value.
In figure \ref{fig:forcedists}, we show the distributions of the monotonic
changes of the external force $\Delta F$ in the steady state, for different
values of the various parameters. Power law scaling of the distributions
consistent with the exponent value $\tau_{\Delta F} = 4/3$ is observed,
with a cut-off of the distributions in agreement with equation 
(\ref{eq:cutoff}). The different $\Delta F$-values appear to be 
uncorrelated in time.

\section{Conclusions}\label{concl}

In this paper we have studied the dynamics of a single particle 
interacting with a cloud of diffusing impurities. In the absence
of external forces, the problem can be mapped to Brownian motion
in a potential within the harmonic approximation of the attractive
particle-impurity interaction. Thus, the monotonic excursions of
the particle are distributed as a power law, 
$P(\Delta x) \sim (\Delta x)^{-\tau_{\Delta x}}$, with 
$\tau_{\Delta x} = 4/3$. For a particle driven with a small
constant velocity (such that the particle is dragging the impurity
cloud without escaping from it), the external force fluctuations
follow the same dynamics, which makes it possible to derive
the probability distribution for the monotonic changes of the
external force scaling as 
$P(\Delta F) \sim (\Delta F)^{-\tau_{\Delta F}}$, again with 
$\tau_{\Delta F} = 4/3$.

While a typical experimentally relevant scenario would correspond
to either a higher dimensional object such as a flexible line 
or a large number of interacting particles interacting with mobile 
impurities, the simple setup of the present study serves as a
convenient starting point for such considerations illustrating the 
relevant phenomena in a transparent manner. One interesting 
observation is that already at the level of a single particle 
interacting with one or more mobile impurities the dynamics has 
scale free features, arising from the properties of simple 
random walks. Physical situations in which these kind
of considerations could be relevant include dislocations interacting
with solute atoms, a subject that has recently attracted considerable 
attention \cite{WAN-00,RIC-03,DEO-05}. In many of these studies,
more realistic interaction forces between dislocations and solute 
atoms have been used, but the focus has been on different quantities 
such as the {\it average} velocity of the dislocation. 

The most intriguing phenomena associated with the presence of
diffusing impurities are the collective effects arising from
the simultaneous interaction of large number of entities with 
each other and with mobile impurities. An example of such a 
system is provided by interacting dislocation ensembles interacting
with diffusing solute atoms in solid solutions, giving rise
to phenomena such as the Portevin-Le Chatelier effect. In the
PLC effect, large numbers of dislocations synchronize their motion
to form macroscopic deformation bands of various kinds. As this is
widely believed to be due to the dynamic interaction of the 
dislocations with the diffusing solute atoms, a natural future
line of research would be to study such effects by considering 
numerically the dynamics of a large number of interacting 
dislocations interacting with diffusing impurities. One further
motivation for such studies could be the recent observation that 
purely stochastic effects can induce switching between collective
motion states \cite{KOL-07}.

% \noindent {\bf Acknowledgment} \\
\ack
LL and MJA gratefully thank the financial support of the European
Commissions NEST Pathfinder programme TRIGS under contract
NEST-2005-PATH-COM-043386. They also acknowledges the financial
support from The Center of Excellence program of the Academy of
Finland.

%\bibliographystyle{apsrev}
%\begin{thebibliography}{99}

\end{document}